\documentclass[11pt]{article}

\usepackage[totalheight = 23cm, totalwidth = 17cm]{geometry}
\usepackage{amssymb,amsmath,amsfonts,amsbsy,graphicx,wrapfig}

\def\mathbi#1{\textbf{\em #1}}

\newcommand{\mpl}{m_{\rm Pl}}
\newcommand{\fnl}{f_{\rm NL}}

\newcommand{\calF}{{\cal F}}

\newcommand{\calR}{{\cal R}}
\newcommand{\calT}{{\cal T}}

\begin{document}

\renewcommand{\theequation}{\arabic{section}.\arabic{equation}}

\begin{titlepage}

\rightline{\footnotesize{CERN-PH-TH/2011-222}} \vspace{-0.2cm}
\rightline{\footnotesize{CPHT-RR 055.0711}} \vspace{-0.2cm}
\rightline{\footnotesize{LPTENS-11-27}} \vspace{-0.2cm}

\begin{center}

\vskip 1.0 cm

{\LARGE  \bf  Effective theories of single field inflation \\ when heavy fields matter}

\vskip 1.0cm

{\large
Ana Ach\'ucarro$^{a,b}$, \hspace{0.2cm} Jinn-Ouk Gong$^{c}$, \hspace{0.2cm} Sjoerd Hardeman$^{a}$,
\\
Gonzalo A. Palma$^{d}$ and Subodh P. Patil$^{e,f}$
}

\vskip 0.5cm

{\it
$^{a}$Instituut-Lorentz for Theoretical Physics, Universiteit Leiden, \mbox{2333 CA Leiden, The Netherlands}
\\
$^{b}$Department of Theoretical Physics, University of the Basque Country,
 48080 Bilbao, Spain
\\
$^{c}$Theory Division, CERN, CH-1211 Gen\`eve 23, Switzerland
\\
$^{d}$Physics Department, FCFM, Universidad de Chile, \mbox{Blanco Encalada 2008, Santiago, Chile}
\\
$^{e}$Centre de Physique Th\'eorique, Ecole Polytechnique and CNRS,
Palaiseau cedex 91128, France
\\
$^{f}$Laboratoire de Physique Th\'eorique, Ecole Normale Superi\'eure,
Paris 75005, France
}

\vskip 1.2cm

\end{center}

\begin{abstract}
We compute the low energy effective field theory (EFT) expansion for single-field inflationary models that descend from a parent theory containing multiple other scalar fields. By assuming that all other degrees of freedom in the parent theory are sufficiently massive relative to the inflaton, it is possible to derive an EFT valid to arbitrary order in perturbations, provided certain generalized adiabaticity conditions are respected. These conditions permit a consistent low energy EFT description even when the inflaton deviates off its adiabatic minimum along its slowly rolling trajectory. By generalizing the formalism that identifies the adiabatic mode with the Goldstone boson of this spontaneously broken time translational symmetry prior to the integration of the heavy fields, we show that this invariance of the parent theory dictates the entire non-perturbative structure of the descendent EFT. The couplings of this theory can be written entirely in terms of the reduced speed of sound of adiabatic perturbations. The resulting operator expansion is distinguishable from that of other scenarios, such as standard single inflation or DBI inflation. In particular, we re-derive how certain operators can become transiently strongly coupled along the inflaton trajectory, consistent with slow-roll and the validity of the EFT expansion, imprinting features in the primordial power spectrum, and we deduce the relevant cubic operators that imply distinct signatures in the primordial bispectrum which may soon be constrained by observations.
\end{abstract}

\vskip 3cm
\begin{quote}
\small
\it{
We dedicate this paper to the memory of our dear colleague and friend, Sjoerd Hardeman. His ideas, insights and diligence permeate every aspect of this work.}
\end{quote}

\end{titlepage}

\newpage
\thispagestyle{empty}
\pagebreak
{\linespread{0.75}\tableofcontents}

\newpage
\setcounter{page}{1}

\section{Introduction}

The application of effective field theory (EFT) techniques to the study of cosmic inflation, constitutes a powerful and systematic approach through which we can calibrate and account for our ignorance of the laws of physics relevant during the very early universe~\cite{Cheung:2007st, Weinberg:2008hq} (see also Ref.~\cite{cliff1}). Without a precise knowledge of the ultraviolet (UV) completion in which the inflaton and its most relevant decay products (which must at least contain the standard model) are embedded, the EFT approach allows us to study the generic predictions of such an embedding on low energy observables. These predictions take the form of specific form factors and self couplings that weight the terms in the derivative expansion of our EFT for all light degrees of freedom. Each set of these couplings represents an equivalence class of UV complete theories which flow to the given effective description valid below the energy scale parametrized by the cutoff $M$.

While it is hard to conceive that one could ever uniquely identify the fundamental theory that gave rise to inflation through the veil of cosmological observations, it is reasonable to suppose that in parametrizing our ignorance through an EFT approach, one can substantially narrow down some of its essential features. 
With the current and upcoming observations on both the cosmic microwave background~\cite{CMB} and large scale structure~\cite{LSS} promising to provide unprecedentedly accurate data, it is important to understand how these might probe and constrain the relevance of any higher dimensional operators during inflation. As a preface to this, it is useful to revisit the conditions under which such operators can become relevant (if even transiently) during effectively single field inflation, consistent with slow-roll, the consistency of the EFT expansion and the presence of the very hierarchy that separates the scale of inflation from the heavy modes that one has putatively integrated out~\cite{Heavy-integration, Achucarro:2010da, Achucarro:2010jv}.

Consider that the derivative expansion of the EFT is an operator expansion weighted by powers of $M$, the scale which defines the separation between heavy and light degrees of freedom. Were we privy to the parent theory before we've integrated out these heavy modes, we would notice that it is articulated in terms of additional mass scales, which will subsequently manifest in the low energy EFT through the couplings that parametrize the derivative expansion. Examples of such additional scales could include the Riemann curvature of the target space of the parent theory, as well as those curvature scales that one can derive from the potential  and its second derivatives which determine the dynamics of the inflaton. A particularly pertinent example of the latter is given by $\kappa$, the radius of curvature in field space of the inflaton trajectory in the parent theory. In constructions representative of inflation realized in a string or a supergravity theory, these additional scales can readily approach the cutoff scale $M$~\cite{sugrainf}. 
Therefore it is a relevant question to ask how the low energy theory accounts for these additional scales in the couplings that weight the derivative expansion, in a manner that is consistent with the very hierarchy that defines the validity of the derivative expansion. In other words, how do the parameters of the parent theory manifest in the EFT expansion? The most straightforward way to answer this
question, is of course to actually integrate out the heavy modes in the parent theory and study the resulting EFT. However in doing so, one encounters several subtleties.

In the specific context of inflationary cosmology, our accounting of
the low energy EFT of inflation has to remain valid over ranges that
accomplish the full 60 odd $e$-folds that is minimally required of
inflation. This is a very tall ask, and is perhaps {\it the} challenge
for inflationary model building today~\cite{challenges}. In many
cases, this could translate into requiring that the EFT we write down
remain valid over very large excursions in field space (in units of
the Planck mass $\mpl$)~\cite{Lyth:1998xn}. Trans-Planckian concerns
notwithstanding~\cite{tp}, in many circumstances, from the perspective
of the parent theory we are forced to account for the other scales
present in the system, which turn out to manifest in the low energy
EFT in an interesting manner.

If for example in the context of single field inflation, were the inflaton trajectory to traverse a sudden\footnote{\label{turnD} In a sense that we shall make precise, if the turn is too ``sudden'', the EFT ceases to be valid~\cite{Achucarro:2010da,Shiu:2011qw,Cespedes:2012hu}. By ``turn'' we specifically mean that the trajectory is not auto-parallel with respect to the sigma model metric, and is parametrized by the radius of curvature in field space $\kappa$, defined as $\kappa^{-1} = V_{N}/\big(\dot\phi_a\dot\phi^a\big)$, where $V_{N}$ is the derivative of the potential normal to the background trajectory, evaluated on the trajectory.} enough turn in field space, one finds that the inflaton will be forced off its adiabatic minimum, in particular cases without even interrupting slow-roll. This results in the transient strong coupling of certain operators in the EFT that
are sensitive to the curvature radius of the trajectory $\kappa$, and which
reduces the effective speed of sound for the long wavelength quanta of the adiabatic mode \cite{Achucarro:2010da, Achucarro:2010jv,Tolley:2009fg,Chen:2009zp}. In addition to enhancing the primordial bispectrum, this transient strong coupling can imprint features in the power spectrum depending on the specific nature of the trajectory\footnote{In Refs.~\cite{Achucarro:2010da, Achucarro:2010jv}, we
  found that the features in the power spectrum were induced by the
  time variation of the reduced speed of sound $\dot c_s$, whereas as
  we shall see shortly, the enhancement of the bispectrum depends on both $c_s$ and $\dot c_s$, where $c_s^{-2} = 1  + 4\dot\phi^2/\big(\kappa^2M^2\big)$, and $\kappa$ is the radius of curvature of the background inflaton trajectory.}. The latter phenomenon arises only after having mindfully integrated out the fast
modes of the theory instead of having truncated them outright. How much the trajectory deviates off the adiabatic minimum is a function of the mass of the heavy fields which couple to the inflaton, the inflaton velocity (which must remain consistent with slow-roll), and  the radius of curvature, $\kappa$, of the trajectory in field space.

Our goal in this paper is to elucidate the general structure, to all orders in perturbations, of the EFT for the adiabatic mode of a theory described by a single effective scalar minimally coupled to gravity at low energies -- a situation that might be dynamically preferred~\cite{Battefeld:2012qx}. In doing so, we have to be mindful of the fact that notions of adiabaticity have to be generalized to account for the fact that the motion of the inflaton through field space can readily force it off its adiabatic minimum, without interrupting slow-roll, even as the scale of inflation remains far below the scale of heavy physics $M$, and in spite of which a low energy EFT description is available and describes the physics of the light modes to the appropriate accuracy.

We will show that the couplings of the EFT are parametrized entirely in terms of the reduced speed of sound $c_s$, which we compute in terms of the parameters of the parent theory. Qualifying the expectation that $c_s$ can at most be a slowly varying function of time~\cite{Cheung:2007st}, or that it cannot become too small without violating perturbative unitarity~\cite{Baumann:2011su}, we find that it is entirely consistent to have {\it transiently} small speeds of sound induced by {\it transient} strong couplings of various operators in the EFT expansion. These are readily generated by turns in the parent theory that cause the background inflaton to deviate off its adiabatic minimum. In spite of this, we are still able to write down an effective theory that captures the relevant physics of the adiabatic mode with the appropriate accuracy, which forces us to consider a generalized notion of adiabaticity that determines the validity of our EFT. This generalization is mandated by the fact that there are two distinct ways in which the low energy description accounts for deviations of the inflaton trajectory off its adiabatic minimum in the parent theory. These can be classed from the perspective of the parent theory, as being due to either:
\begin{itemize}
\item 
\emph{Strong turns}, where the adjective refers to the degree to which the speed of sound $c_s$ is reduced, which is commensurate with how far off the adiabatic minimum of the potential the background inflaton trajectory is forced by its evolution, or 
\item 
\emph{Sudden turns}, where the adjective refers to the rate of change of the speed of sound $\dot c_s$, induced by changes in the radius of curvature of the trajectory.
\end{itemize}
From now on, all references to \emph{strong} and \emph{sudden} will reference these two definitions.
Each of these has a different scope for being realized consistently in the low energy description. Sustained strong turns ($c_s \ll 1, \dot c_s \sim 0$ over several $e$-folds of expansion) are difficult to maintain in tandem with a consistent derivative expansion (without invoking additional special symmetries)~\cite{Baumann:2011su}. However within certain limits that we will elaborate upon, sudden turns that aren't too strong ($c_s \lesssim 1, \dot c_s \sim c_sH$ over a fraction of an $e$-fold) can readily and consistently generate transient strong (but not too strong) couplings in the EFT~\cite{Achucarro:2010da} (see also Refs.~\cite{Shiu:2011qw,Cespedes:2012hu}). In addition, the nature of the turn thus classified results in identifiably distinct contributions to  observables: a strong turn for example, can reduce the speed of sound of the adiabatic mode, but for a constant radius of curvature, serves to only renormalize the power spectrum whilst generating equilateral non-Gaussianity~\cite{Achucarro:2010da,Chen:2009zp,Chen:2009we}. Sudden turns on the other hand can generate features in the power spectrum~\cite{Achucarro:2010da} as well various non-trivial shapes in the bispectrum~\cite{us}.
The positive detection of these observables would allow us to infer non-trivial information about the EFT in which inflation is embedded\footnote{See also Ref.~\cite{heavy-fields} for a complimentary discussion of the effects of heavy fields relevant at the beginning of inflation.}.

\subsection{Overview}

Although multi-field inflation models have been thoroughly studied
during the past decade, it has only recently been realized that multi-field scenarios in which the
masses of extra fields remain much larger than the rate of expansion
possess a non-trivial single-field limit~\cite{ Achucarro:2010da, Achucarro:2010jv, Tolley:2009fg, Cremonini-etal} beyond the
standard slow-roll paradigm\footnote{
  See Refs.~\cite{Chen:2009zp,Chen:2009we} for a
  study of when the masses of the extra fields approach the scale of
  inflation, set by $H$, the Hubble factor during inflation. 
  See also Ref.~\cite{multiEFTothers} for an EFT approach to 
  multi-field inflation when the extra fields are light.}. In this
limit, high frequency modes (set by some mass scale $M\gg H$) are
irrelevant and any heavy fields act as spectators that only effect the
low energy dynamics by inducing non-trivial higher dimensional
interactions for the inflaton, appropriately weighted by powers of
$M$. These higher dimensional operators have a variety of consequences
such as reducing the speed of sound of adiabatic
perturbations~\cite{Achucarro:2010jv,Tolley:2009fg}, generating large
levels of primordial non-Gaussianity in the limit $M\to H$~\cite{Chen:2009zp,Chen:2009we}, in addition to generating features in the power spectrum that pass the threshold of being potentially observable~\cite{Achucarro:2010da}.

In what follows we attempt to further our understanding of the
structure of the single field effective theory derived from parent
theories with a single light inflaton among multiple other fields with
masses much larger than the scale of inflation. By analyzing the terms
of the derivative expansion that emerge from integrating out heavy
degrees of freedom, we show that the speed of sound $c_s$ for the
adiabatic perturbations emerges as \emph{the} salient book-keeping
device around which successive terms in the derivative expansion
naturally organize themselves. This reduced speed of sound will be a
functional of the background inflaton trajectory, itself dependent on
the field geometry and the multi-field potential.

Specifically, we implement the integration of heavy fields through a generalization of the framework introduced in Ref.~\cite{nicolis} to study cosmological perturbations in general FRW backgrounds (applied to inflation in Ref.~\cite{Cheung:2007st}), where the Goldstone boson corresponding to broken time translational invariance is identified with the adiabatic mode of our system, now consisting of our multi-field parent theory minimally coupled to gravity. To pare down the discussion to its physical essentials, we consider the parent theory to consist of a canonical two-field system, where a hierarchy of scales between the masses of the two fields persists along the entire inflationary trajectory\footnote{For an investigation for the case of two fields inflation with a weak hierarchy of masses, see Refs.~\cite{2-field,Peterson:2010np}.}. 
As we discuss shortly, the treatment that follows straightforwardly extends to more general parametrizations of the parent theory. By explicitly integrating out the heavy modes, we are able to deduce an effective theory valid to all orders in field perturbations, allowing us to identify the operators which become relevant enough to imprint on the cosmological observables within the threshold of observational sensitivity. It is worth emphasizing a few key points before we commence our analysis:

\begin{itemize}

\item As a corollary of diffeomorphism invariance, the parent multi-field theory possesses a time translational invariance that is broken on general cosmological backgrounds. It follows that all fluctuations around this background that can be gauged away by a local time reparametrization-- the so called adiabatic modes-- can thus be related to the Nambu-Goldstone boson (or Goldstone boson for short) corresponding to this broken symmetry. 

\item Given that the action for the Goldstone boson is forced to be invariant under shift symmetries~\cite{GT}, one can identify the structure of the derivative expansion in the resulting effective theory. Specifically, the requirement that constant solutions exist to all orders (even after having integrated out the heavy fields that couple to the Goldstone mode), constrains the terms that can appear in its derivative expansion. This is the legacy of the broken time translation invariance, passed down from the parent theory to the EFT. The Goldstone boson of broken time translational invariance of the parent theory, in conforming with convention is denoted\footnote{As homage to its particle physics origins, wherein the pion is realized as the pseudo-Goldstone boson of spontaneously broken chiral flavor symmetry in QCD.} $\pi = \pi(t,\mathbi{x})$, and is readily identified with the one introduced in Ref.~\cite{nicolis} and specifically applied to inflation in Ref.~\cite{Cheung:2007st}, after we integrate out the heavy fields that couple to it.

\item The relevant quantity which naturally organizes the various terms in the EFT expansion order by order turns out to be the speed of sound $c_s$ of the adiabatic mode. After integrating out the heavy modes, all relevant couplings can be expressed in terms of $c_s$, in a manner that can be distinguished from other theories whose primary feature is a reduced speed of sound, such as DBI inflation~\cite{DBI} or general $P(X)$ theories~\cite{px}.

\item Contrary to the naive expectation that suggests that in the slow-roll limit $( -\dot H / H^2 \to 0)$ these couplings must vary slowly in time, we show that their variation can be considerably large if the inflaton trajectory deviates from a geodesic (i.e. turns) sufficiently strongly or suddenly. This allows for certain operators in the EFT expansion that one might have previously neglected to become transiently strongly coupled, consistent with perturbative unitarity and the consistency of our derivative expansion, in a way that can imprint features on the primordial power spectrum~\cite{Achucarro:2010da} due to the induced variations of the speed of sound. In particular, it is easy to infer from the EFT that sufficiently rapid variation of these couplings will necessarily introduce cubic derivative interactions for the Goldstone mode, implying correlated non-Gaussian signatures in the primordial bispectrum~\cite{us}.

\end{itemize}

The above realizations straightforwardly extend beyond the parametrization of the parent theory examined in this study, so long as the original theory respects diffeomorphism invariance. In particular, although we have chosen to work with a canonical two-scalar field model for the benefit of clarity, our treatment is carried out in a geometrically covariant language which allows our results to easy extend to more general setups involving more than two fields or a non-trivial sigma model metric: through field redefinitions, one can always work in a basis where one transforms away the sigma model metric at the expense of introducing derivative interactions (and corrections to the potential), which are easily accounted for.

We organize our investigation as follows: in Section~\ref{Minkowski} we discuss the simple example of a two-field model in Minkowski spacetime, where the scalar fields break time translation by acquiring time-dependent expectation values. This example exhibits in an uncomplicated setting, all of the essential features that are found in the more complicated setup where gravity must be accounted for. In particular, we will show how time translation symmetry of the full theory imposes strong constraints on the non-perturbative structure of the EFT deduced by integrating out the heavy degree of freedom. Then, in Section~\ref{sec:multi-field} we move on to study the more involved case of two scalar fields minimally coupled to gravity. There, we derive the action for multi-field perturbations in two different gauges: The comoving gauge, where the adiabatic perturbation is readily expressed in terms of the gauge invariant field $\calR$, and the flat gauge, where the Goldstone boson $\pi$ takes the leading role in describing adiabatic perturbations. In Section~\ref{sec:EFT} we explicitly integrate out the heavy field and deduce the EFTs in both gauges. In Section~\ref{sec:Discussion} we proceed to discuss the validity of the resulting effective theories, and examine some of their salient features. We compare our results with previous work and discuss its ramifications for future observations, after which we offer our concluding remarks in Section~\ref{sec:Conclusions}.

\section{A toy model on Minkowski space-time}  
\label{Minkowski}
\setcounter{equation}{0}

It turns out that most of the essential features encountered in the context of inflationary cosmology are already present in the much simpler setup of scalar field dynamics on Minkowski spacetime. Although such a setup ignores gravity, many of the structures inherent to perturbation theory around inflationary backgrounds are due to symmetries that are already present on Minkowski spacetimes. Thus for conceptual clarity, we warm up with this example before coupling our system to gravity. But our analysis is straightforwardly extendable to include additional fields. Consider the following action
\begin{equation}\label{original-action-minkowski}
S = \int d^4x \left[ -\frac{1}{2}\eta^{\mu\nu}\partial_\mu\phi^a\partial_\nu\phi_a - V(\phi) \right] \, ,
\end{equation}
where the $a$-index labels the two fields. This action has the property of being invariant under space-time translations $x^{\mu} \to \widehat{x}^\mu = x^\mu + \xi^\mu$, where $\xi^\mu$ is a constant vector. In other words, by defining $\widehat{\phi}^a(\widehat{x}) = \phi^a(x)$, we trivially recover the same action, but with $\widehat{\phi}^a(\widehat{x})$ in place of $\phi^{a}(x)$. The equations of motion that result from (\ref{original-action-minkowski}) are
\begin{equation}
\ddot \phi^a - \Delta\phi^a + V^a = 0 ,
\end{equation}
where $\Delta \equiv \delta^{ij} \partial_i \partial_j$. There is a particular set of background solutions $\phi^a(t,\mathbi{x}) =  \phi_0^a(t)$ that preserve spatial translations but spontaneously break the time translation symmetry of the original action. These solutions are homogeneous and isotropic in the reference frame given by $x^{\mu} = (t,\mathbi{x})$, and satisfy the following equation of motion:
\begin{equation}\label{minkowski-eq-of-motion-background}
\ddot \phi_0^a (t) + V^a (\phi_0) = 0 \, .
\end{equation}
We now consider a few corollaries resulting from the time translational symmetry of the original action. Assuming that we have a solution to (\ref{minkowski-eq-of-motion-background}) of the form $\phi_0^a (t)$ with given boundary conditions, then because the invariance of the action under finite time translations $t \to \widehat{t} = t + \xi^{0}$, one can generate a family of non-trivial solutions $\phi_{\Delta}^a (t)$ out of $ \phi_0^a (t)$ as
\begin{equation}
\phi_{\Delta}^a (t) =   \phi_0^a (t + \Delta \mathcal T)  \, ,
\end{equation}
where $\Delta \mathcal T$ is a constant of our choice. Of course, a similar relation must hold for the initial conditions, fixed at time $t_0$: $\phi^a_{\Delta} (t_0) = \phi^a_0 (t_0 + \Delta \mathcal T)$ and $\dot \phi^a_{\Delta} (t_0) = \dot \phi^a_0 (t_0 + \Delta \mathcal T)$. In other words, while it is true that $\phi_0$ spontaneously breaks time translational invariance, the generator of the broken symmetry can be used to produce a one parameter family of solutions $ \phi_\Delta^a (t) $ from the base solution $ \phi_0^a (t) $. This seemingly inane observation turns out to have far reaching consequences for the perturbative expansion of the system.

\subsection{Background} 
\label{Sec: Background-Minkowski}

Before jumping into the intricacies of perturbation theory, we briefly review
some useful geometrical background identities and fix our notation. A given background solution $ \phi_0^a (t) $ defines a trajectory in field space parameterized by time $t$. It is therefore natural to define the tangent vector $T^a = T^a(t)$ to the trajectory as
\begin{equation}\label{tdef}
T^a \equiv \frac{\dot\phi^a_0}{\dot \phi_0} \, ,
\end{equation}
where $\dot \phi_0 = \sqrt{ \dot \phi^a \dot \phi_a}$ is the rapidity of the
scalar field's vacuum expectation value. Differentiating $\dot \phi_0^2 = \dot \phi^a \dot \phi_a$ with respect to time, we get $ 2 \dot \phi_0
\ddot \phi_0 = 2 \dot \phi_a \ddot \phi^a = - 2 \dot \phi_0 T_a V^a$,
from which one obtains the projection of equation of motion of the scalar field (\ref{minkowski-eq-of-motion-background}) along the trajectory
\begin{equation}\label{eq-of-motion-parallel-mink}
\ddot \phi_0 + V_T = 0 \, ,
\end{equation}
where we use the notation $V_T \equiv T^a V_a$. Since we are dealing with a two-field system, we may also define a unit vector $N^a = N^a(t)$ normal
to the trajectory, {\it i.e.} $N^a T_a = 0$. To define it in a unique
manner, it is enough to write the normal as $N^a = \epsilon^{a b} T_b$ where
$\epsilon_{a b}$ is the totally antisymmetric $2 \times 2$ matrix with
unit entries. This choice fixes the orientation of $N^a$ with respect
to the trajectory. Given the pair $T^a$ and $N^a$, we may now \emph{define}
an angular velocity $\dot \theta$ for the trajectory as
\begin{equation}\label{def-alpha}
\dot\theta = - N_a \dot T^a \, .
\end{equation}
Because we work in a
two-dimensional field space, it is easy to show that
\begin{equation}\label{T and N -  alpha}
\dot T^a = -\dot\theta N^a \, , \qquad \dot N^a = \dot \theta T^a \, .
\end{equation}
Moreover, by projecting the equation of motion
(\ref{minkowski-eq-of-motion-background}) along $N^a$, it is possible to derive an identity determining $\dot \theta$ in terms of the slope of the potential $V_N \equiv N^a V_a$ along the normal direction~\cite{Achucarro:2010da,Achucarro:2010jv}
\begin{equation}\label{relation-alpha-V-N}
\dot \theta =  \frac{V_N}{\dot\phi_0} \, .
\end{equation}
This tells us that as soon as the trajectory is subject to a turn,
the background scalar field deviates off the adiabatic minimum of the potential, defined by $V_N = 0$. The relative sign between $\dot\theta$ and $V_N$ follows the geometrical intuition
that the turning trajectory will be displaced onto the outer slope of the valley of the potential whenever a turn takes place. The background parameter $\dot
\theta$ will play an important role in what follows.

\subsection{Perturbations}

\begin{wrapfigure}{T}{0.5\textwidth}
\vspace{-10pt}
  \begin{center}
    \includegraphics[width=0.48\textwidth]{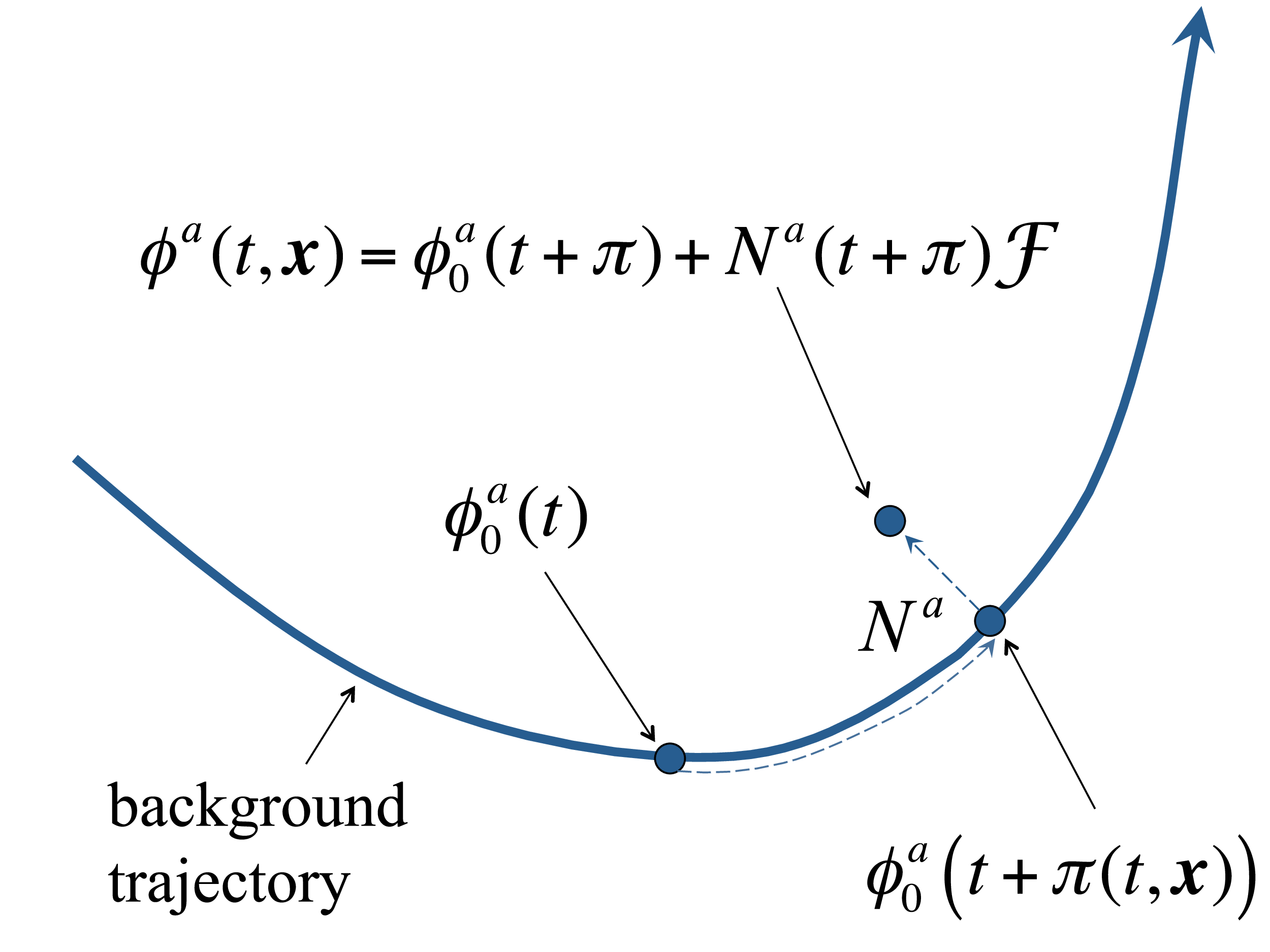}
  \end{center}
\end{wrapfigure}

We now consider perturbations around the background solution $\phi_0^a(t)$. We begin by fixing some definitions: we define the fields $\pi = \pi(t,\mathbi{x})$ and $\mathcal F = \mathcal F (t,\mathbi{x})$ to represent departures from the homogeneous and isotropic background solution $\phi_0^a(t)$ as
\begin{equation}\label{def-mink-pi-F}
\phi^a(t,\mathbi{x}) =  \phi_0^a (t + \pi  ) + N^a (t + \pi ) \mathcal F \, ,
\end{equation}
where $N^a (t + \pi )$ stands for the vector $N^{a}$ evaluated at $t + \pi$. Notice that $\pi$ represents deviations from $\phi_0^a(t)$ exactly along the path defined by the background solution, whereas $\mathcal F$ parametrizes deviations off the trajectory, but evaluated at $t + \pi$. In fact, the field $\mathcal F$ lives in the tangent space $\mathbb{T}_{\phi (t + \pi)}$ spanned by $N^a (t + \pi )$.

Before determining the action for the perturbations in terms of $\pi$ and $\mathcal F$, we can already anticipate certain features of it to all orders in perturbation theory. Since $\phi_\Delta^a(t) = \phi_0^a(t + \Delta \mathcal T)$ with $\Delta \mathcal T$ an arbitrary constant is also a solution of the system (provided that the initial conditions are set appropriately), it follows that there must be a non-trivial solution for the set of fields  $\pi$, $\mathcal F$ such that
\begin{equation}\label{particular-non-trivial-solution-mink}
\pi = {\rm constant} \qquad {\rm and} 	\qquad \mathcal F = 0 \, .
\end{equation}
This can be done by setting $\pi$ to be constant everywhere on some initial hypersurface with $\dot\pi \equiv \dot\calF \equiv \calF \equiv 0$. Given the existence of this solution, along with the spatial and rotational invariance of the background solution, one finds that the interactions of the theory written in terms of $\pi$ and $\mathcal F$ are greatly constrained. For example, we can anticipate that the Lagrangian for $\pi$ and $\mathcal F$ is allowed to contain cubic interactions of the form $\dot \pi^3$, $\dot \pi^2 \pi$, $\pi^2 \Delta \pi$,  $\dot \pi \Delta \pi^2$, $\pi \dot \pi \mathcal F$ and $\dot \pi^2 \mathcal F$, but cannot contain interactions proportional to $\pi^3$,  $\pi^2 \dot \pi$ and $\pi^2 \mathcal F$ simply because the latter do not lead to equations of motions for which (\ref{particular-non-trivial-solution-mink}) is a solution. Moreover, for the particular case of long wavelength perturbations, where spatial derivatives of $\pi$ can be ignored, we can immediately infer that one of the two linearly independent solutions to the equations of motion must be $\pi =$ constant.

A further corollary that follows from the existence of solution
(\ref{particular-non-trivial-solution-mink}) is relevant for when we eventually integrate out $\mathcal F$. If we linearize the equation of motion for $\mathcal
F$ but keep $\pi$ to all orders, on general grounds it will be of the
form: 
\begin{equation}
\mathcal O_\calF\left(\partial_t^2, \Delta, \pi\right) \mathcal F = \Pi \left(\pi, \dot \pi , \Delta\pi\right) \, ,
\end{equation}
where ${\mathcal O}_{\mathcal F} \left(\partial_t, \Delta, \pi\right) $ is a general linear
operator acting on $\mathcal F$ which may include couplings that
depend on $\pi$. On the other hand, $\Pi \left(\pi, \dot \pi , \Delta
\pi\right)$ is a function of $\pi$ and space-time derivatives of it, such
that it vanishes when $\pi = \,$constant. This is because the
configuration $\pi = \,$constant and $\mathcal F = 0$ must be a
solution of the system.  Therefore, if we define $M^2(\pi)$ as the result
of letting $\mathcal O_{\mathcal F} \left(\partial_t, \Delta, \pi \right)$ act
on a constant,
\begin{equation}
\mathcal O_{\mathcal F} \left(\partial_t^2, \Delta, \pi \right) \times {\rm constant} 
= M^2(\pi) \times {\rm constant} \, ,
\end{equation}
the integration
of $\mathcal F$ can be symbolically expressed (with the provision of appropriate boundary conditions) as 
\begin{equation}
\mathcal F_{\pi} = \mathcal O^{-1}_{\mathcal F}
\left(\partial_t^2, \Delta, \pi\right) \Pi \left(\pi, \dot \pi , \Delta\pi\right) =
\left[ \frac{1}{M^2(\pi)} + \cdots \right] \Pi \left(\pi, \dot \pi ,
\Delta\pi\right) \, ,
\end{equation}
where dots represents powers of linear operators
suppressed by higher order powers of $M^{2}$. This expansion is only convergent at energy scales much smaller than $M$, which therefore defines the cutoff scale for the resulting EFT. That is, $\mathcal F$ will contribute to the effective action of $\pi$ operators as a succession of higher dimensional operators suppressed in powers of $M^2$, and with a spacetime derivative structure consistent with $\pi = \,$constant as a non-trivial solution. Consequently, the resulting EFT for $\pi$ will inherit the property that $\pi = \,$constant is a solution of the system, due to the original symmetry of the parent multi-field theory.

\subsection{Action for $\pi$ and $\mathcal F$}

We now verify the assertions of the previous section by direct computation of the action for the perturbation fields $\pi$ and $\mathcal F$. Substituting the decomposition (\ref{def-mink-pi-F}) into (\ref{Sec: Background-Minkowski}), and using the identities (\ref{tdef}) - (\ref{T and N - alpha}), we find that the action becomes
\begin{equation}\label{action-mink-pi-F}
S  =   \frac{1}{2} \int \!\! d^4 x \bigg\{   \left(  \dot \phi_0   + \dot \theta \calF \right)^2  \left[ 1 + 2 \dot \pi + \dot \pi^2 - (\nabla\pi)^2 \right] + \dot \calF^2 - (\nabla\calF)^2  - 2 V(\phi) \bigg\} \, ,
\end{equation}
where every background quantity (for instance $\dot \phi_0$ and $\dot \theta$) is evaluated at $t + \pi$, and where $\phi$ appearing in $V (\phi) $ stands for $\phi \to  \phi_0^a (t + \pi ) + N^a (t + \pi ) \mathcal F $. It is important to realize that the action in (\ref{action-mink-pi-F}) is exact in the sense that we have only assumed that we know the background solution $\phi_0^a(t)$ and have used it to parametrize the theory in a more suitable way for the class of phenomena we are interested in studying. Observe that terms that are excluded by the existence of the constant solution to all orders, such as $\dot \pi \pi^2$ or $\pi^3$ do indeed appear in the action when expanded in powers of $\pi$ and $\mathcal F$, seemingly contradicting the results of the previous section. However these terms will only ever appear in combinations which drop out of the action, and we may rewrite the theory in a way that the previous property is manifest. For this, it is useful to expand the theory in powers of $\mathcal F$, in which case it takes the form
\begin{align}\label{action-mink-pi-F-2}
S  =&   \frac{1}{2} \int \!\! d^4 x \bigg\{    \dot \phi_0^2  \left[   \dot \pi^2 - (\nabla\pi)^2 \right]  +    2 \dot \theta \dot \phi_0 \mathcal F    \left[  2 \dot \pi + \dot \pi^2 - (\nabla\pi)^2 \right]
\nonumber\\
&
\hspace{1.3cm} + \dot \theta^2 \calF^2   \left[  2 \dot \pi + \dot \pi^2 - (\nabla\pi)^2 \right]   + \dot \calF^2 - (\nabla\calF)^2  - M_{\rm eff}^2  \calF^2 - \frac{1}{3} V_{NNN} \calF^3  + \cdots  \bigg\}
\nonumber\\
& +  \frac{1}{2} \int d^4 x \left[ \dot \phi_0^2 (t + \pi ) \left( 1 + 2\dot \pi \right) - 2 V \big( \phi_0 (t + \pi )   \big) \right] \, ,
\end{align}
where the ellipsis denotes higher order terms coming from the expansion of the potential about the configuration $\mathcal F = 0$, where all background quantities are evaluated at $t + \pi$, and where
\begin{equation}\label{def-effective-mass-mink}
M_{\rm eff}^2 \equiv V_{NN} - \dot \theta^2 \, ,
\qquad
V_{NN} \equiv N^a N^b V_{a b} \, ,
\qquad
V_{NNN} \equiv N^a N^b N^c V_{a b c} \, . 
\end{equation}
Notice that $V_{NN}$, which is the projection of $V_{ab}$ along the direction $N^a$, corresponds to the bare mass of fluctuations normal to the trajectory. However during a turn, the effective potential for the heavy fluctuation $\mathcal F$ receives a centripetal correction, reflected by the appearance of the term $- \dot \theta^2$ at the right hand side of the definition of the effective mass $M_{\rm eff}^2$. Given that all background quantities are on shell, the terms of the last line of (\ref{action-mink-pi-F-2}) group into a total derivative and so can be discarded. Now we see explicitly that, in agreement with our previous analysis, the action is consistent with solutions of the form written in (\ref{particular-non-trivial-solution-mink}).

\subsection{Effective action for $\pi$}
\label{sec: effective-action-mink}

If the field $\calF$ is sufficiently massive, we can derive an effective theory for $\pi$ obtained by integrating out $\calF$. Such a theory will contain all possible terms consistent with the requirement that the constant solution exists, and consistent with spatial translational and rotational invariance, i.e. the residual symmetries of broken time translational invariance. To see this, we first note that the equation of motion for $\calF$ is given by
\begin{equation}\label{eq-of-motion-F-mink}
\ddot \calF - \Delta\calF + \left\{ M_{\rm eff}^2 -   \dot \theta^2  \left[  2 \dot \pi + \dot \pi^2 - (\nabla\pi)^2 \right]   \right\}  \calF 
-  \dot \theta  \dot \phi_0 \left[  2 \dot \pi + \dot \pi^2 - (\partial \pi)^2 \right]  + V_{NNN} \calF^2 + \cdots = 0 \, , 
\end{equation}
where again the ellipsis stands for higher order contributions coming from the potential expansion. To integrate out $\mathcal F$, we solve for it to {\it linear} order and plug the formal solution back into the action wherever $\mathcal F$ appears\footnote{It can be shown \cite{us} that this prescription exactly reproduces the tree level effective action. Loop contributions, a very interesting story in their own right, are not captured by this prescription. We will return to these contributions in a future study.} and expand in powers of $M^{2}_{\rm eff}$, which to first order means disregarding space-time derivatives acting on $\mathcal F$. This gives us an operator relation determining $\mathcal F$ in terms of $\pi$. We denote the first order solution as $\mathcal F_{\pi}$, given by
\begin{equation}\label{F-effective}
\mathcal F_{\pi}  = \frac{ \dot \theta (t + \pi )   \dot \phi_0 (t + \pi ) \left[  2 \dot \pi + \dot \pi^2 - (\nabla\pi)^2 \right]  }{ M_{\rm eff}^2 (t + \pi  )   - \dot \theta^2 (t + \pi ) \left[ 2 \dot \pi + \dot \pi^2 - (\nabla\pi)^2 \right]  } \, .
\end{equation}
The effective action that results is then:
\begin{equation}\label{eq:action-eft-minkowski}
S  =   \frac{1}{2} \int d^4 x  \dot \phi_0^2 (t + \pi )  \left\{    \dot \pi^2 - (\partial \pi)^2 +   \frac{(c_s^{-2}  - 1) \left[  2 \dot \pi + \dot \pi^2 - (\nabla \pi)^2 \right]^2  }{ 4  - (c_s^{-2} - 1) \left[  2 \dot \pi + \dot \pi^2 - (\nabla \pi)^2 \right]  }   \right\} \, , 
\end{equation}
where $c_s$ is the speed of sound of $\pi$ perturbations defined by
\begin{equation}\label{def-speed-of-sound}
c_s^{-2} \equiv 1 + \frac{4 \dot \theta^2}{M_{\rm eff}^2} \, .
\end{equation}
We have neglected higher order terms in the above as we are only interested in the low energy effective theory for the Goldstone mode $\pi$ to leading order in $M^{-2}_{\rm eff}$  (for example, the term proportional to $V_{NNN}$ would have resulted in a contribution $\sim M^{-6}_{\rm eff}$). Notice that there are only two background quantities defining the couplings of the theory, namely the overall factor $ \dot \phi_0^2 (t) $ and $c_s(t)$, both evaluated at $t + \pi$. Because this effective theory was deduced by neglecting derivatives of $\calF$  in (\ref{eq-of-motion-F-mink}), its validity hinges on the following adiabaticity conditions:
\begin{equation}\label{condition-1}
\big| \ddot \calF _{\pi} \big|  \ll M_{\rm eff}^2  \left| \calF _{\pi} \right| \, ,  
\qquad
\left| \Delta\calF_{\pi} \right|  \ll M_{\rm eff}^2   \left| \calF_{\pi} \right|  \,  .
\end{equation}
These conditions represent what is meant by working within the low energy regime, and more interestingly, they also imply restrictions on how fast background quantities such as $\dot \theta$, $\dot \phi_0$ and $M_{\rm eff}^2$ are allowed to vary in time without invalidating the effective theory. Furthermore, in expanding the theory to third order in spacetime derivatives of $\pi$, we obtain
\begin{align}\label{effective-action-mink}
S  =&   \frac{1}{2} \int d^4 x  \dot \phi_0^2   \bigg\{ c_s^{-2}  \dot \pi^2 - (\nabla \pi)^2
+  \left(\frac{1}{c_s^2} - 1\right)  \dot \pi  \left[  \dot \pi^2 - (\nabla \pi)^2  \right]
+ \left(\frac{1}{c_s^2} -1\right)^2  \frac{ \dot \pi^3   }{ 2    }
\nonumber\\
&
\hspace{1.8cm} + 2 \frac{\ddot \phi_0}{\dot \phi_0}  \left[ \frac{\dot\pi^2}{c_s^2} - (\nabla \pi)^2 \right] \pi - 2 \frac{\dot c_s}{c_s^3}   \dot \pi^2  \pi
   \bigg\} \, ,
\end{align}
where background quantities are now evaluated at $t$. We find that all possible terms consistent with the symmetries discussed previously are indeed present, and arise from couplings in the parent theory involving $\mathcal F$. The last two terms in (\ref{effective-action-mink}) come from the expansion of background quantities about the configuration  $\pi=0$.  If both $\dot \phi_0^2$ and $c_s$ vary slowly in time, we may safely disregard those terms. In this case, the effective action deduced coincides with the general effective action deduced in Ref.~\cite{Cheung:2007st} for the inflationary case. We will elaborate upon this connection in more detail in Section~\ref{sec:multi-field}. Furthermore, we note in passing that operators weighted by higher derivatives of $c_s$ would appear at higher orders in our expansion in $M_{\rm eff}^{-2}$

\subsection{On the origin of the reduced speed of sound}

We immediately infer from (\ref{def-speed-of-sound}) that as soon as the scalar field trajectory is subject to a turn, the speed of sound is reduced. Before considering the ramifications of this during inflationary cosmology (and coupling these scalar fields to gravity), we briefly reflect upon the reduced speed of sound in the effective theory obtained from having integrated out $\mathcal F$. For this, it is convenient to consider the particular case where all the background quantities, including $\dot \theta$, remain constant. This situation corresponds to a circular trajectory as for instance could be the case for a ``mexican hat" potential. This example may be solved exactly to quadratic order in the parent theory (for details, consult Ref.~\cite{Achucarro:2010jv}) giving the following schematic solutions in terms of the eigenmodes $(\pi_-,\mathcal F_-)$ and $(\pi_+ , \mathcal F_+)$:
\begin{align}
\pi_c =& \pi_+ e^{i \omega_+ t}  + \pi_- e^{i \omega_- t} \, ,
\\
\mathcal F =&   \mathcal F_+ e^{i \omega_+ t}  + \mathcal F_- e^{i \omega_- t} \, ,
\end{align}
where $\pi_c = \dot \phi_0 \pi / c_s$ is the canonically normalized perturbation along the path, and the two frequencies $\omega_{\pm}$ in terms of the wavenumber $k$ are given by
\begin{equation}\label{efreq}
2 \, \omega^{2}_{\pm} = M_{\rm eff}^2 c_s^{-2} + 2 k^2   \pm M_{\rm eff}^2 c_s^{-2}  \sqrt{  1 +  \frac{4 k^2}{M_{\rm eff}^2 c_s^{-2}} ( 1 - c_s^{2})   } \, .
\end{equation}
Crucially, whenever $\dot \theta = 0$ one has $\pi_+ = \mathcal F_- = 0$ and the high frequency modes $\omega_+^2 = M^2 + k^2$ are exclusively associated with $\mathcal F$ and the low frequency modes $\omega_-^2 = k^2$ are exclusively associated with $\pi_c$. That is $\calF$ and $\pi$ are aligned with the heavy and light eigenmodes respectively. However, if the trajectory undergoes a turn ($\dot \theta \neq 0$), $\calF$ and $\pi$ no longer align with the appropriate eigenmodes. Integrating out the heavy mode essentially consists in integrating out the degree of freedom with the larger eigenfrequencies $\omega_{+}$. From (\ref{efreq}) we see that to even have such a hierarchy we require
\begin{equation}\label{condition-2}
c_s^2k^2 \ll M_{\rm eff}^2
\end{equation}
to be satisfied. Notice that condition (\ref{condition-2}) is a refined version of condition (\ref{condition-1}) for the particular case where the background quantities $\dot \phi_0^2$, $M_{\rm eff}^2$ and $\dot \theta^2$  are constant. In this regime one has $\omega_- = c_s  k$ as a dispersion relation for long wavelength excitations, from where one indeed infers the effective speed of sound for the adiabatic modes as $c_s^{-2} = 1 + 4 \dot \theta^2/M_{\rm eff}^2$. 

Thus we see that the reduced speed of sound does not arise from the dynamics of any fast modes in the parent system {\it per se} (which are appropriately decoupled), but rather due to the mixing induced by the turn, which induces a non-zero projection onto the mode $\mathcal F_-$ around the turn, which oscillates in phase with the adiabatic modes at low frequencies thus affecting the speed at which $\pi_c$ propagate.

\section{Multi-field inflation with a single light field}
\label{sec:multi-field}
\setcounter{equation}{0}

We now consider inflationary models, where gravity has to be incorporated into the picture. Gravity does not play a significant role in the dynamics of the heavy fields (with $M_{\rm eff}\gg H$), and so the basic features inferred of the structure of the effective theory for the adiabatic mode persist. In part, this is because heavy fields are not associated to any symmetries of the background space-time as opposed is the case with the Goldstone mode $\pi$. We begin by considering the action
\begin{align}\label{original-action-inflation}
S = \int d^4 x \sqrt{-g} \left[ \frac{\mpl^2}{2} R - \frac{1}{2} g^{\mu\nu}\partial_{\mu}\phi^a\partial_{\mu}\phi_a - V(\phi^a) \right] \, .
\end{align}
It is convenient to use the Arnowitt-Deser-Misner (ADM) formulation~\cite{Arnowitt:1962hi} to write the metric as
\begin{equation}\label{ADM-metric}
ds^2 = -N^2dt^2 + h_{ij} \left( dx^i+N^idt \right) \left( dx^j+N^jdt \right) \, ,
\end{equation}
where $N$, $N^i$ and $h_{ij}$ are the lapse function, shift vector and the metric on the spatial slices foliating space-time. Then, the pure gravity part of the action reads
\begin{equation}\label{ADM-action}
S^\text{(G)} = \frac{\mpl^2}{2} \int d^4x N\sqrt{h}  \left[ R^{(3)} + \frac{1}{N^2} \left( E^{ij}E_{ij} - E^2 \right) \right] \, ,
\end{equation}
where $ R^{(3)} $ is the spatial Ricci scalar constructed from $h_{ij}$ and $E_{ij} \equiv \big( \dot{h}_{ij} - h_{jk}N^k{}_{|i} - h_{ik}N^k{}_{|j} \big)/2$, with a vertical bar denoting a covariant derivative with respect to $h_{ij}$.

We now retrace the steps we carried out in Section~\ref{Minkowski}. To begin with, the scalar field equations of motion deduced from (\ref{original-action-inflation}) are given by
\begin{equation}
-\Box\phi^a + V^a = 0 \, ,
\end{equation}
where  $\Box \equiv \partial_\mu\left(\sqrt{-g}g^{\mu\nu}\partial_\nu\right)/\sqrt{-g}$.
We consider homogeneous and isotropic solutions of this system, characterized by a background metric of the form
\begin{equation}
ds^2 = -dt^2 + a^2(t)\delta_{ij}dx^idx^j \, ,
\end{equation}
where $a(t)$ is the scale factor. Homogeneous and isotropic scalar field solutions $\phi_0^a(t)$ satisfy the equations of motion
\begin{equation}\label{scalar-field-eom-inflation}
\ddot\phi_0^a(t) + 3H\dot\phi_0^a(t) + V^a(\phi_0) = 0 \, ,
\end{equation}
where $H \equiv \dot{a}/a$ is the Hubble parameter. The previous equations may be supplemented with the Friedmann equation which relates $H$ with the energy density through the relation
\begin{equation}\label{Friedmann}
3\mpl^2H^2 = \frac{1}{2}\dot\phi_0^2 + V (\phi_0) \, ,
\end{equation}
where $\dot \phi_0^2 = \dot \phi^a \dot \phi_a$. Another useful identity is obtained by differentiating (\ref{Friedmann}) with respect to time:
\begin{equation}
\dot{H} = - \frac{\dot\phi_0^2}{2\mpl^2} \, .
\end{equation}
Similar to the Minkowski space example studied in Section~\ref{Minkowski}, although this set of solutions spontaneously breaks the time translational invariance of the original theory, we can use it to deduce some relevant properties of the perturbed theory. Let us assume that we already have a set of solutions $ \phi_0^a (t) $ and $a(t)$ with given boundary conditions. Then, because the original theory (\ref{original-action-inflation}) is invariant under time translations, we can generate a family of solutions out of the background quantities $\phi_0^a(t)$ and $a(t)$, of the form
\begin{align}
\phi^a_{\Delta}(t) \equiv & \phi_0^a(t+\Delta\calT) \, ,
\\
a_{\Delta}(t) \equiv & a(t+\Delta\calT) e^{ \Delta\calR }\, ,
\end{align}
where $\Delta\calT$ and $\Delta\calR$ are arbitrary constants of our choice. These constants will play the same role in allowing us to deduce the general non-perturbative structure of couplings for the fields describing deviations from homogeneity and isotropy.

\subsection{Parametrizing perturbations}

With only a few minor modifications, we can employ many of the geometrical identities used to parameterize the background trajectory of the scalar fields in Minkowski space-time. For instance, just as before, we can define tangential and normal unit vectors to the trajectory as $T^a = \dot\phi_0^a/\phi_0$ and $N^a = \epsilon^{ab}T_b$, respectively. Then projecting (\ref{scalar-field-eom-inflation}) along the tangential direction $T^a$, we obtain
\begin{equation}
\ddot\phi_0 + 3H\dot\phi_0 + V_T = 0 \, ,
\end{equation}
where $V_T \equiv T^aV_a$. In addition, we may use the same definition for $\dot \theta$ given in (\ref{def-alpha}) to denote the angular velocity of the curvilinear inflaton trajectory, and the relations expressed in (\ref{T and N - alpha}) persist, with $\dot \theta$ given by\footnote{An alternative notation to describe turns consists of defining the dimensionless parameter $\eta_{\perp} = \dot \theta/ H$, which was originally introduced in Ref.~\cite{etaperp}, and its notation stems from the definition of a vector $\eta^{a} \equiv -\ddot\phi_0^a/(H\dot\phi_0)$~\cite{Gong:2002cx} in order to extend the definition of the usual $\eta$ parameter to the multi-field case. In this setting, $\eta_\perp$ is simply the projection of $\eta^a$ along the normal direction $N^a$. We should emphasize however that $\eta_\perp$ should not be regarded as a slow-roll parameter, and that it may acquire large values without implying departure of slow-roll inflation, in the sense that $H$ may continue to evolve adiabatically~\cite{Achucarro:2010da,Peterson:2010np}.}
\begin{equation}
\dot \theta = \frac{V_N}{\dot\phi_0} \, ,
\end{equation}
where $V_N \equiv N^aV_a$. We now proceed to parameterize departures of this system from homogeneity and isotropy. For the matter sector of the theory we proceed exactly as in (\ref{def-mink-pi-F}), i.e. 
\begin{equation}\label{def-inf-pi-F}
\phi^a(t,\mathbi{x}) = \phi_0^a(t+\pi) + N^a(t+\pi)\calF \, .
\end{equation}
For the gravitational sector however, we need to proceed a bit more carefully. In the ADM decomposition (\ref{ADM-metric}), $N$ and $N^i$ are Lagrangian multipliers while $h_{ij}$ is dynamical. Thus, to define perturbations within the gravity sector, we write $h_{ij}$ as
\begin{equation}\label{pert-spatial-metric}
h_{ij} = a(t+\pi)e^{2\calR}\gamma_{ij} \, ,
\end{equation}
where $\det\gamma=1$. To simplify our analysis, we disregard vector and tensor degrees of freedom, and fix the linear spatial gauge by setting $\gamma_{ij} = \delta_{ij}$~\cite{Noh:2003yg}. We note that $\pi$ relates to field fluctuations along its trajectory, $\calF$ to displacements away from the trajectory (which translate into isocurvature perturbations), and $\calR$ to the spatial curvature perturbation of a spatial foliation parametrized by time $t + \pi$.

Just as we did in Minkowski space-time, we can immediately infer some non-perturbative properties of the action for the fields $\pi$, $\calR$ and $\calF$. Since $\phi_\Delta^a(t) = \phi_0^a(t+\Delta\calT)$ and $a_{\Delta}(t) = a(t+\Delta\calT) e^{\Delta\calR}$ are also solutions of the system with $\Delta\calT$ and $\Delta\calR$ arbitrary constants, there must exist a non-trivial solution for $\pi$, $\calR$ and $\calF$ such that\footnote{Notice that, in the absence of $\calF$, the argument presented here is similar to the one introduced in the $\delta N$ formalism~\cite{Gong:2002cx,deltaN} to prove that $\calR$ is conserved in the long wavelength limit.}
\begin{align}
\label{particular-non-trivial-solution-inf1}
\pi = & {\rm constant} \, ,
\\
\label{particular-non-trivial-solution-inf2}
\calR = & {\rm constant} \, ,
\\
\label{particular-non-trivial-solution-inf3}
\calF = & 0 \, ,
\end{align}
in addition with the following requirement on the lapse and shift functions:
\begin{align}
\label{particular-non-trivial-solution-inf-21}
N = & 1 \, ,
\\
\label{particular-non-trivial-solution-inf-22}
N^i = & 0 \, .
\end{align}
This implies that many terms in the perturbative expansion are immediately excluded. In particular, since $N$ and $N^i$ are Lagrange multipliers, by solving their constraint equations one must obtain a set of constraints for which (\ref{particular-non-trivial-solution-inf-21}) (\ref{particular-non-trivial-solution-inf-22}) are automatically satisfied for the configuration (\ref{particular-non-trivial-solution-inf1}), (\ref{particular-non-trivial-solution-inf2}) and (\ref{particular-non-trivial-solution-inf3}). Furthermore, provided that $\calF$ is sufficiently massive, integrating it out will necessarily result in a theory where $\pi = $constant and $\calR =$constant are preserved as non-trivial solutions to all orders in perturbation theory.

\subsection{Gauge transformation properties}

Given that we will eventually be interested in integrating out $\calF$, it is instructive to examine the manner in which $\pi$ and $\mathcal R$ are interrelated by gauge transformations. For this, consider a general coordinate transformation of the form:
\begin{equation}
x^\mu \to \widehat{x}^\mu = x^\mu + \xi^\mu \, .
\end{equation}
For clarity, in this discussion we consider the particular case of a single field. It is useful then to expand (\ref{def-inf-pi-F}) and (\ref{pert-spatial-metric}) in terms of $\pi$ and $\mathcal R$, and compare the results with the variables encountered in the more conventional treatments of cosmological perturbation theory~\cite{cpt}. More precisely, by defining $\delta \phi (t,x)  \equiv  \phi(t,\mathbi{x}) - \phi_0(t) $ and $\varphi$ through the relation $h_{ij}(t,\mathbi{x})  \equiv a^2(t) (1 + 2\varphi) \delta_{ij} $, we find
\begin{align}
\delta\phi(t,\mathbi{x}) = & \dot\phi_0\pi + \frac{\ddot\phi_0}{2}\pi^2 + \cdots \, ,
\\
\varphi (t,\mathbi{x}) = &    H\pi + \calR + \left( H^2 + \frac{\dot{H}}{2} \right)\pi^2 + 2H\pi\calR + \calR^2  + \cdots  \, .
\end{align}
We start by examining the gauge transformations at linear order in the perturbations. To begin with, notice that we have already set $\xi^i=0$ at linear order by demanding the condition $\gamma_{ij}=\delta_{ij}$, so that $\widehat{x}^i=x^i$.  Then, under the remaining coordinate transformation $t \to
\widehat{t} = t+\xi^0$, we see that $\delta\phi=\dot\phi_0\pi$ and
$\varphi=H\pi+\calR$ transform as
\begin{align}
\label{deltaphixform_goldstone}
\widehat{\delta\phi} = & \dot\phi_0\left( \pi - \xi^0 \right) \, ,
\\
\label{gammaijxform_goldstone}
\widehat{\varphi} = & H \left( \pi - \xi^0 \right) + \calR \, .
\end{align}
This shows that $\widehat{\pi}=\pi-\xi^0$, so that $\pi$ can indeed be regarded as the Goldstone boson associated with the spontaneous breaking of time translational symmetry. We furthermore see that $\widehat {\mathcal R} = \mathcal R$.

We can now choose various foliations, represented by $\widehat{x}^\mu$, of differing physical significance. For example, we may choose to work on comoving slices, where $\widehat{\delta\phi} = 0$. From (\ref{deltaphixform_goldstone}) we see that this is equivalent to set $\xi^0 = \pi$. This in turns implies, from (\ref{gammaijxform_goldstone}), that $\widehat \varphi $ coincides with the gauge invariant spatial curvature perturbation as $\varphi_{\delta\phi} = \calR$. 
Thus any property which our {\em original} $\calR$ possesses is inherent on the comoving slice.
On the other hand, we may choose to work on a flat slicing where the spatial metric is unperturbed, i.e. $\varphi = 0$. Then, from (\ref{gammaijxform_goldstone}) we find that it is necessary to set $\xi^0 = \pi + \calR/H$, implying that 
\begin{equation}
\widehat{\pi} = -\frac{\calR}{H} \, .
\end{equation}
However, the cost of choosing this gauge is that now the non-perturbative structure of the theory describing $\widehat{\pi}$ does not share the same properties as the original $\pi$. That is, the theory in the flat gauge does not admit $\widehat{\pi}=\text{constant}$ as a solution, but instead admits the solution $\widehat{\pi} = \text{constant}/H$. In the slow-roll regime, where $H$ is almost a constant, this is not so significant.

It is now straightforward to examine the gauge transformation properties of $\pi$ and $\mathcal R$ to any desired order in the perturbations. Let us examine explicitly the situation at second order. There we find:
\begin{equation}\label{deltaphi_pi_xform}
\widehat{\delta\phi} = \dot\phi_0\left( \pi-\xi^0 \right) + \frac{\ddot\phi_0}{2} \left( \pi-\xi^0 \right)^2 - \dot\phi_0 \left( \dot \pi- \dot \xi^0 \right) \xi^0 \, .
\end{equation}
Thus, we see that setting $\xi^0 = \pi$ gives $\widehat{\delta\phi} = 0$ up to second order without any additional conditions. The gauge invariant curvature perturbation $\varphi_{\delta\phi}$ is then given by
\begin{equation}\label{gaugeinvR}
\varphi_{\delta\phi} = \calR + \calR^2 - \dot\calR\pi - \frac{1}{2a^2} \left[ \psi^{,i}\pi_{,i} + \frac{1}{2}\pi^{,i}\pi_{,i} - \Delta^{-1} \left( \psi_{,i}\pi_{,j} + \frac{1}{2}\pi_{,i}\pi_{,j} \right)^{,ij} \right] \, ,
\end{equation}
where we have defined $N_i = \partial_i \psi$. 
(\ref{gaugeinvR}) is gauge invariant and may thus be evaluated on any generic slice. For instance, if we evaluate it on a comoving slice, where $\pi=0$, we simply have $1+2\varphi_{\delta\phi}=e^{2\calR}$ up to second order. In fact, in the comoving gauge the equivalence between $\varphi_{\delta\phi}$ and $\calR$ holds non-perturbatively, as can be immediately seen from (\ref{def-inf-pi-F}) and (\ref{pert-spatial-metric}) by simply setting $\pi=0$. However this trivial relation does not hold in the flat gauge, i.e. the simple relation $\calR = -H\pi$ is not valid beyond linear order. This result indicates a special significance to the comoving gauge.

\subsection{Action in two gauges}

We now deduce the action governing the dynamics of perturbations in two different gauges, which will turn out to be useful when studying the effective theory resulting from integrating out $\calF$.

\subsubsection{Comoving gauge: $\calR$ and $\calF$}

We first consider the action for the fields $\calR$ and $\calF$ in comoving gauge, defined by $\pi = 0$. Here we may define the field content from the very beginning as:
\begin{align}
\phi^a(t,\mathbi{x}) = & \phi_0^a(t) + N^a(t)\calF(t,\mathbi{x}) \, ,
\\
h_{ij}(t,\mathbi{x}) = & a^2(t) e^{2\calR(t,\mathbi{x})} \delta_{ij} \, .
\end{align}
Introducing these definitions back into the original action (\ref{original-action-inflation}) we derive the action to all orders in $\calR$ and $\calF$ as
\begin{align}
\label{total-action-R-F-previous}
S = & \int d^4x Na^3 \frac{e^{3\calR}}{2} \left\{ -2\mpl^2\frac{e^{-2\calR}}{a^2} \left[ 2\Delta\calR + \left(\nabla\calR\right)^2 \right] + \frac{\mpl^2}{2N^2} \left( N^i{}_{,j}N^j{}_{,i} + \delta_{ij}N^{i,k}N^j{}_{,k} - 2N^i{}_{,i}N^j{}_{,j} \right) \right.
\nonumber\\
& \hspace{2.7cm} - \frac{6\mpl^2}{N^2}\left( H + \dot\calR - N^i\calR_{,i} \right)^2 + \frac{4\mpl^2}{N^2} \left( H + \dot\calR - N^i\calR_{,i} \right) N^j{}_{,j}
\nonumber\\
& \hspace{2.7cm} \left. + \frac{1}{N^2} \left( \dot\phi_0 +\dot \theta \calF \right)^2 + \frac{1}{N^2} \left( \dot\calR - N^i\calF_{,i} \right)^2 - h^{ij}\calF_{,i}\calF_{,j} - 2V\left(\phi_0^a+N^a\calF\right) \right\} \, .
\end{align}
Similar to the situation encountered in (\ref{action-mink-pi-F}), it is not obvious from (\ref{total-action-R-F-previous}) that $\calR =$constant and $\calF=0$ constitute a non-trivial solution of the system to all orders in perturbation theory. To reach a form of the action where these solutions are manifest, it is convenient to expand it in powers of $\calF$, and up to total derivatives rewrite the action as
\begin{align}\label{total-action-R-F}
S = & \int d^4x Na^3 \frac{e^{3\calR}}{2} \left\{ -2\mpl^2\frac{e^{-2\calR}}{a^2} \left[ 2\Delta\calR + \left(\nabla\calR\right)^2 \right] + \frac{\mpl^2}{2N^2} \left( N^i{}_{,j}N^j{}_{,i} + \delta_{ij}N^{i,k}N^j{}_{,k} - 2N^i{}_{,i}N^j{}_{,j} \right) \right.
\nonumber\\
& - \frac{12\mpl^2H}{N} \left( \frac{1}{N}-1 \right)\dot\calR - \frac{1}{N} \left( \frac{1}{N}+N-2 \right) \left( 6\mpl^2H^2-\dot\phi_0^2 \right)
\nonumber\\
& + \frac{\mpl^2}{N^2} \left[ 12HN^i\calR_{,i} - 6\left( \dot\calR-N^i\calR_{,i} \right)^2 + 4 \left( H + \dot\calR - N^i\calR_{,i} \right)N^j{}_{,j} \right] + \frac{1}{N^2} \left( \dot\calF - N^i\calF_{,i} \right)^2
\nonumber\\
& \left. - h^{ij}\calF_{,i}\calF_{,j} + \dot \theta \left( \frac{1}{N^2}-1\right) \left( 2\dot\phi_0\calF + \dot \theta \calF^2 \right) - M_{\rm eff}^2\calF^2 - \frac{1}{3}V_{NNN}\calF^3 + \cdots \right\} \, ,
\end{align}
where, as before, $M_{\rm eff}^2 \equiv V_{NN}-\dot \theta^2$, $V_{NN} \equiv N^aN^bV_{ab}$ and $V_{NNN} \equiv N^aN^bN^cV_{abc}$.
The only terms that could potentially spoil the existence of solutions of the form (\ref{particular-non-trivial-solution-inf-21}) and (\ref{particular-non-trivial-solution-inf-22}) are in the first two terms of the second line. They are however, linear and quadratic in terms of $\delta N \equiv N - 1$ and therefore, after solving the constraint equations, both $N$ and $N^i$ necessarily become functions of $\calR$ and $\calF$ in such a way that $N=1$ and $N^i = 0$, when $\calR =$constant and $\calF=0$. This may easily be verified by solving the constraint equations at linear order. If we write $N_i = \partial_i \psi + \widetilde{N}_i$, where $\widetilde{N}_i$ is a pure vector satisfying $\partial_i \widetilde{N}^i=0$, then we find the linear solutions as
\begin{align}
\label{linear-constraints-R1}
N  = & 1 + \frac{\dot\calR}{H} \, ,
\\
\label{linear-constraints-R2}
\frac{\Delta}{a^2}\chi = & \epsilon\dot\calR - \frac{  \dot \theta  \dot\phi_0}{H \mpl^2}   \calF \, ,
\\
\label{linear-constraints-R3}
\widetilde{N}^i  = & 0 \, ,
\end{align}
where $\epsilon \equiv - \dot H / H^2$ and  $\chi \equiv \psi+\calR/H$. From here we see that indeed $N=1$ and $N^i = 0$ when $\calR = $constant and $\calF = 0$.

\subsubsection{Flat gauge: $\pi$ and $\calF$}

As discussed perviously, we may choose to work in the flat gauge $\varphi = 0$ where $\pi$ becomes the dynamical field of interest, but at the expense of losing the property that $\pi = $constant and $\mathcal F = 0$ are solutions of the system. It has been understood however that there are a few advantages of adopting this gauge~\cite{Cheung:2007st}, particularly when the inflationary system is in a regime where the mixing between gravity and the Goldstone boson $\pi$ can be neglected. In this gauge the scalar field and the 3-metric $h_{ij}$ take the form
\begin{align}
\phi^a(t,\mathbi{x}) = & \phi_0^a(t+\pi) + N^a(t+\pi)\calF \, ,
\\
h_{ij} = & a^2(t)\delta_{ij} \, .
\end{align}
By inserting these two expression back into the original action (\ref{original-action-inflation}), we obtain
\begin{align}\label{action-Goldstone-full}
S = & \int d^4x \frac{Na^3}{2} \left\{ -\frac{6\mpl^2H^2}{N^2} + \frac{4\mpl^2 H}{N^2}N^i{}_{,i} + \frac{\mpl^2}{2N^2} \left( N^i{}_{,j}N^j{}_{,i} + \delta_{ij}N^{i,k}N^j{}_{,k} - 2N^i{}_{,i}N^j{}_{,j} \right) \right.
\nonumber\\
& \hspace{2cm} + \frac{1}{N^2} \left( \dot\phi_0+ \dot \theta \calF \right)^2 \left[ \left( 1 + \dot\pi - N^i\pi_{,i} \right)^2 - \frac{N^2}{a^2}\left(\nabla\pi\right)^2 \right]
\nonumber\\
& \hspace{2cm} \left. + \frac{1}{N^2} \left( \dot\calF - N^i\calF_{,i} \right)^2 - \frac{\left(\nabla\calF\right)^2}{a^2} - 2V\left(\phi_0^a+N^a\calF\right) \right\} \, ,
\end{align}
where $a$ and $H$ are being evaluated at $t$, whereas every other background quantity is being evaluated at evaluated at $t+\pi$ instead of $t$. The linear solutions of the constraint equations are found to be
\begin{align}
\label{linear-constraints-pi1}
N = & 1 + H\epsilon\pi \, ,
\\
\label{linear-constraints-pi2}
\frac{\Delta}{a^2}\psi = & -H\epsilon \left( \dot\pi - H\epsilon\pi \right) - \frac{\dot \theta \dot\phi_0}{H \mpl^2}  \calF \, .
\end{align}
We are now ready to discuss the derivation of the effective theory for the adiabatic mode.

\section{Effective single field theories}
\label{sec:EFT}
\setcounter{equation}{0}

In this section we deduce the effective single field theory resulting from integrating out the heavy field $\calF$ in the two previously discussed gauges. As in our discussion of Section~\ref{sec: effective-action-mink}, we identify the speed of sound $c_s$ of adiabatic mode as $c_s^{-2} = 1 + 4\dot \theta^2 / M_{\rm eff}^2 $.

\subsection{Effective action for $\calR$}

We wish to compute the effective action for $\calR$ by integrating out $\calF$. We begin by deriving the equation of motion for $\calF$:
\begin{align}
& \frac{d}{dt} \left[ \frac{a^3e^{3\calR}}{N} \left( \dot\calF - N^i\calF_{,i} \right) \right] - \partial_i \left[ \frac{a^3e^{3\calR}}{N}\left( \dot\calF - N^j\calF_{,j} \right)N^i + h^{ij}\calF_{,j} \right]
\nonumber\\
& = Na^3e^{3\calR} \left\{ \left( \frac{1}{N^2}-1 \right) \dot\phi_0 \dot \theta - \left[ M_{\rm eff}^2 - \left( \frac{1}{N^2}-1 \right)\calF \right] + \cdots \right\} \, .
\end{align}
If we ignore the kinetic terms of $\calF$ and solve the resulting algebraic equation for $\calF$, we find:
\begin{equation}\label{F-R}
\calF_\calR = \frac{\dot\phi_0 \dot \theta \left(N^{-2}-1\right)}{M_{\rm eff}^2-\dot \theta^2 \left(N^{-2}-1\right)} .
\end{equation}
This expression gives us the low energy dependence of $\calF$ in terms of $N$. Of course, after solving the constraint equation for $N$ one obtains $\calF$ in terms of $\mathcal R$. For instance, to linear order in the perturbations, after using the constraint equation (\ref{linear-constraints-R1}) for $N$, we find that:
\begin{equation}
\calF_\calR = -\frac{2\dot\phi_0 \dot \theta }{H M_{\rm eff}^2}\dot\calR . \,
\end{equation}
Note that we may now compute the effective action for $\mathcal R$ and the Lagrange multipliers $N$ and $N^i$, by inserting (\ref{F-R}) into the general action (\ref{total-action-R-F}) after only quadratic terms in $\mathcal F$ are considered and disregarding its kinetic terms. This prescription is accurate up to loop contributions, to leading order in $M^{-2}_{\rm eff}$. For simplicity, let us compute the action up to third order in perturbation, which require us to solve constraint equations to linear order as (\ref{linear-constraints-R1}), (\ref{linear-constraints-R2}) and (\ref{linear-constraints-R3}). Observe that to linear order $N$ is not affected by $\calF$.
Then we find the linear solution $\chi$ in terms of $\calR$ as
\begin{equation}
\label{chi-linear}
\frac{\Delta}{a^2}\chi = \left( 1 + \frac{4 \dot \theta^2}{M_{\rm eff}^2} \right)\epsilon\dot\calR \equiv \frac{\epsilon}{c_s^2}\dot\calR \, ,
\end{equation}
where we have identified $c_s^{-2} = 1 + 4\dot \theta^2 / M_{\rm eff}^2 $.
Substituting this result back into (\ref{total-action-R-F}) and expanding up to third order in $\calR$ gives us the desired EFT for $\calR$, which we write as $S = S_2 + S_3$ with:
\begin{align}
\label{action-R-S2}
S_2 = & \int d^4x \frac{a^3\epsilon\mpl^2}{c_s^2} \left[ \dot\calR^2 - c_s^2\frac{(\nabla\calR)^2}{a^2} \right] \, ,
\\
\label{action-R-S3}
S_3 = & \int d^4x a^3 \left[ -\epsilon\mpl^2\calR\frac{(\nabla\calR)^2}{a^2} + 3\frac{\epsilon\mpl^2}{c_s^2}\dot\calR^2\calR + \epsilon\mpl^2 \frac{\left(1-c_s^2\right)^2 - 2}{2c_s^4} \frac{\dot\calR^3}{H} \right.  
\nonumber\\
&
\hspace{1.5cm} \left.  + \frac{\mpl^2}{2a^4} \left\{ \left( 3\calR - \frac{\dot\calR}{H} \right) \left[ \psi^{,ij}\psi_{,ij} - (\Delta\psi)^2 \right] - 4\calR^{,i}\psi_{,i}\Delta\psi \right\} \right] \, .
\end{align}
Inspection of (\ref{action-R-S2}) shows that, indeed, the speed of sound is given by $c_s^{-2} = 1 + 4\dot \theta^2 / M_{\rm eff}^2 $.

\subsection{Effective action for $\pi$}

We now work out the effective action for $\pi$ in the flat gauge. Following the same steps in integrating out the heavy field $\calF$, we begin with the solution of $\calF$ in terms of $\pi$ as
\begin{align}
\label{F-effective-pi-inf}
\calF_\pi = & \frac{\gamma \dot\phi_0  \dot \theta }{M_{\rm eff}^2-\gamma \dot \theta^2} ,
\\
\gamma \equiv & \frac{1}{N^2}-1 + \frac{1}{N^2} \left[ 2\left(\dot\pi-N^i\pi_{,i}\right) + \left(\dot\pi-N^i\pi_{,i}\right)^2 - N^2\frac{(\nabla\pi)^2}{a^2} \right] \, .
\end{align}
Note that (\ref{F-effective-pi-inf}) agrees with (\ref{F-R}) once we use $\pi = - \calR/H$. As already mentioned, the benefit of expressing the theory in terms of $\pi$ is particularly apparent when discussing the slow-roll regime $\dot{H}\to0$. If we now insert (\ref{F-effective-pi-inf}) back into (\ref{action-Goldstone-full}) and keep only quadratic terms in $\mathcal F$, neglecting kinetic terms (which contribute at next order in $M^{-2}_{\rm eff}$), we obtain the desired effective action for $\pi$. The contribution to the effective action from the matter sector is given by
\begin{align}\label{action-Goldstone-full-2}
S \supset & \int d^4x \frac{Na^3}{2} \left\{ \frac{\dot\phi_0^2}{N^2} \left[ \left( 1 + \dot\pi - N^i\pi_{,i} \right)^2 - \frac{N^2}{a^2}\left(\nabla\pi\right)^2 \right] - 2V + \frac{\gamma^2  \dot\phi_0^2 \dot \theta^2 }{M_{\rm eff}^2-\gamma \dot \theta^2} + \cdots \right\} \, .
\end{align}
In the regime in where gravity is negligible, the contribution from the gravitational sector action can be neglected, and contributions from $N^i$ and $1-N^2$ in the matter sector can be disregarded. The resulting effective action thus becomes
\begin{equation}\label{action-Goldstone-full-3}
S_{\rm eff} = \int d^4x \frac{a^3}{2}\dot\phi_0^2 \left\{ \dot\pi^2 - \frac{(\nabla\pi)^2}{a^2} + \frac{ (c_s^{-2} - 1) \left[ 2\dot\pi + \dot\pi^2 - (\nabla\pi)^2/a^2 \right]^2}{4 - (c_s^{-2} - 1)  \left[ 2\dot\pi + \dot\pi^2 - (\nabla\pi)^2/a^2 \right]} \right\} \, .
\end{equation}
Because we are in a regime where the effects of gravity are neglected, it should comes as no surprise that this action coincides with (\ref{eq:action-eft-minkowski}) obtained on a Minkowski background. To compare this result with the general Goldstone boson action discussed in Ref.~\cite{Cheung:2007st}, it is useful to expand this action to third order in $\pi$ and make use of the background identity $\dot\phi_0^2 = - 2\mpl^2\dot{H}$,
\begin{align}\label{eft-action-pi-inflation}
S_{\rm eff} = & -\int d^4x a^3\mpl^2\dot{H} \left\{ c_s^{-2}  \dot\pi^2 - \frac{(\nabla\pi)^2}{a^2}   + \left(c_s^{-2} - 1 \right)\dot\pi \left[ \dot\pi^2 - \frac{(\nabla\pi)^2}{a^2} \right] + \left( c_s^{-2} - 1 \right)^2 \frac{\dot\pi^3}{2}
\right.
\nonumber\\
& \hspace{3cm} \left.  -2  \frac{\dot c_s}{c_s^3}  \pi\dot\pi^2 - 2 H \eta_\parallel  \pi \left[ c_s^{-2}  \dot\pi^2 - \frac{(\nabla\pi)^2}{a^2} \right]  \right\} \, ,
\end{align}
where $\eta_\parallel \equiv -\ddot\phi_0/(H\dot\phi_0)$.
As argued in Ref.~\cite{Cheung:2007st}, in order to be consistent with the limit $\dot{H} \to 0$, one should neglect the time variation of other background quantities, which amounts to neglecting $\eta_\parallel$ and $\dot c_s$ in (\ref{eft-action-pi-inflation}). It has been shown however, that large time variations of $c_s$ are indeed consistent with adiabatic evolution of $H$~\cite{Achucarro:2010da}, and so in general one should not disregard terms proportional to $\dot c_s$.

\section{Discussion} 
\label{sec:Discussion}
\setcounter{equation}{0}

Equations (\ref{action-R-S3}) and  (\ref{eft-action-pi-inflation}) constitute the cubic order effective actions for $\calR$ and $\pi$ respectively once the heavy field $\calF$ has been integrated out. The conditions determining the validity of these EFT's are analogous to those discussed for the case of Minkowski backgrounds, given by (\ref{condition-1})~and~(\ref{condition-2}). That is, we must demand both $\calF_\calR$ and $\calF_\pi$ of equations (\ref{F-R}) and (\ref{F-effective-pi-inf}) to satisfy
\begin{align}
\label{adiab-condition-1}
\left| \Delta \calF \right|  \ll & M_{\rm eff}^2    \left| \calF  \right| \, ,
\\
\label{adiab-condition-2}
\big| \ddot \calF \big|  \ll & M_{\rm eff}^2  \left| \calF \right|  \, .
\end{align}
These conditions express how quickly the heavy field $\mathcal F$ is allowed to vary without exciting high frequency modes. For instance, the first condition (\ref{adiab-condition-1}) translates into the condition on comoving wavelengths $k$ as
\begin{equation}
 c_s^2 \frac{k^2}{a^2} \ll M_{\rm eff}^2 \, ,
\end{equation}
which is equivalent to condition (\ref{condition-2}). 
However, because the study of perturbations during inflation requires us to consider the stretching of wavelengths from the intermediate regime $H^2 \ll  c_s^2  k^2 / a^2 \ll M_{\rm eff}^2$, where one imposes initial conditions, to the long wavelength regime $c_s^2 k^2 / a^2 \ll H^2$, we may opt to characterize the validity of the EFT in terms of background quantities as
\begin{equation}\label{condition-4}
H^2 \ll M_{\rm eff}^2 \, .
\end{equation}
It is important to recognize that we could have only assumed condition (\ref{adiab-condition-2}) and not neglected spatial derivative contributions to the kernel of $\mathcal F$ when integrating it out. This is easily done at quadratic order~\cite{Achucarro:2010da,Achucarro:2010jv,Cespedes:2012hu}, but in the more general case, would have forced us to deal with more complicated contributions to the effective action (that are not necessarily negligible for short wavelength modes), obscuring the main message of this work. However, we foresee no problem in addressing this issue straightforwardly in principle.

The second condition~(\ref{adiab-condition-2}) is more relevant, as it allows us to assess how quickly background quantities are allowed to vary in time without invalidating the EFT expansion. In particular, one obtains restrictions on how sudden a turn of the inflaton trajectory can be, furnishing the {\it generalized adiabaticity conditions} we have alluded to earlier. Indeed, in Ref.~\cite{Cespedes:2012hu} it was found that (\ref{adiab-condition-2}) leads to the following condition on the variation of the angular velocity $\dot \theta$
\begin{equation}\label{adiabat-cond-angular}
\left| \frac{d}{dt}  \log \dot \theta \right| \ll M_{\rm eff} \, ,
\end{equation}
which asserts that the rate at which the turn is accelerating must be much smaller than $M_{\rm eff}$, which determines the high frequency of heavy fluctuations. In terms of the speed of sound, this condition can be expressed as
\begin{equation}\label{adiabat-cond-speed}
\left| \frac{d}{dt}  \log \left(c_s^{-2} - 1\right) \right|  \ll M_{\rm eff} \, , 
\end{equation}
where we have assumed that any variation of $c_s$ due to the turn is primarily dictated by the time variation of $\dot \theta$. 

Other constraints on the validity of the EFT are based on considerations involving the perturbative validity of the theory. To discuss this, it is useful to compare our result (\ref{action-Goldstone-full-3}) with the EFT action deduced in Ref.~\cite{Cheung:2007st} in terms of the Goldstone boson. There, a set of mass scales $M_n^4$ is introduced which, to cubic order, appears in (\ref{eft-action-pi-inflation}) as:
\begin{equation}\label{EFT-Cheung}
S_{\pi}  = - \int d^4 x a^3    \left\{ \dot H \left[ \dot \pi^2  -    a^{-2}( \nabla \pi)^2 \right]
- 2 M_2^4   \dot \pi  \left[ \dot \pi +  \dot \pi^2 -  \frac{(\nabla \pi)^2}{a^2}   \right]  +    \frac{4 M_3^4}{3}  \dot \pi^3  + \cdots  \right\} \, .
\end{equation}
In this notation, the speed of sound takes the form $c_s^{-2} = 1 - 2 M_2^4 /\big(\mpl^2\dot H\big)$. In Ref.~\cite{Cheung:2007st} the mass scales $M_n^4$ represent the couplings determining the expansion of the EFT action in powers of $1+ g^{00} = 1 - N^{-2}$ in the comoving gauge. To compare their formalism with our results, it is enough to expand the effective action (\ref{action-Goldstone-full-2}) in powers of $- \gamma$ and subsequently identify the coefficients of the expansion with $M_n^4$ as:
\begin{equation}\label{M-n-eft}
M_n^4 = (-1)^{n} n! \,  \big| \dot H \big|  \mpl^2 \left(  \frac{ c_s^{-2} - 1 }{4}  \right)^{n-1} \, .
\end{equation}
That is, to leading order in $M^{-2}_{\rm eff}$ {\it all of the higher dimensional operators of the effective theory are uniquely characterized by the speed of sound $c_s$}. Given that for $c_s < 1$ the perturbative expansion contains non-renormalizable terms, we must be weary of strong coupling effects that might affect the validity of EFT. Indeed, the strong coupling scale $\Lambda$ is determined by $\Lambda^4 \sim 4\pi \mpl^2 \big|\dot H \big| c_s^{5}/\left(1-c_s^2\right)$ (see Refs.~\cite{Cheung:2007st,Baumann:2011su} for a discussion on this) and therefore we may demand that
\begin{equation}\label{bound-strong-coupling}
M_{\rm eff}^4 \lesssim 4\pi \mpl^2 \big|\dot H \big| \frac{c_s^5}{1-c_s^2} \, . 
\end{equation}
For comparison, we notice that in Ref.~\cite{Baumann:2011su} the scale of new physics $\omega_{\rm new}$ coincides with our definition of $M_{\rm eff}$.
If this condition is respected, the EFT is valid all the way up to energies below our reference scale $M_{\rm eff}$, before entering the strong coupling regime. Notice that this condition imposes restrictions on how small $c_s$ may become for a certain fixed value of the effective mass $M_{\rm eff}$. Because the speed of sound already depends on the effective mass (through the relation $c_s^{-2} = 1 + 4 \dot \theta^2 / M_{\rm eff}^2$) we may alternatively write this condition as a constraint between the angular velocity $\dot \theta$ and the $M_{\rm eff}$:
\begin{equation}
4\dot\theta^2M_{\rm eff}^2 \left( 1+\frac{4\dot\theta^2}{M_{\rm eff}^2} \right)^{3/2} \ll 4\pi\mpl^2 \big|\dot{H}\big| \, .
\end{equation}

Also, it is interesting to notice that (\ref{M-n-eft}) represents a specific prediction of the present type of theories that might be contrasted with other models of effective single field inflation. For example, we observe that the mass scales  $M_2^4$ and $M_3^4$ are related as
\begin{equation}\label{eq: M2-M3 multi-field}
\frac{M_3^4}{M_2^4}  = - \frac{3}{4} \left(c_s^{-2}  - 1\right) \, ,
\end{equation}
which may be compared to the result found for DBI inflation, where these scales are related through
\begin{equation}\label{eq: M2-M3 DBI}
\frac{M_3^4}{M_2^4}  =  3  \left( c_s^{-2}  - 1\right) \, .
\end{equation}
Equation (\ref{eq: M2-M3 multi-field}) provides a means of discriminating the class of models we've considered with other models of single field inflation with a lowered speed of sound\footnote{One can show however that the ratio $M_2^4/M_1^4$ for the effective action we derived agrees with that for DBI inflation.}. It gives the main contribution to the bispectrum. If the speed of sound is suppressed, these couplings generically predict~\cite{Achucarro:2010da,Chen:2006nt}
\begin{equation}
\fnl \sim c_s^{-2} \, ,
\end{equation}
where the main contributions come from the operators $\dot \pi (\nabla \pi)^2$ and $\dot \pi^3$ which are precisely combinations of $M_2^4$ and $M_3^4$.

Finally, it is worth emphasizing that the form of the EFT (\ref{EFT-Cheung}) provided in Ref.~\cite{Cheung:2007st} assumes that background quantities vary slowly compared to $H$. From the multi-field point of view, however, this is not the only relevant timescale determining the evolution of the background. In particular, $M_{\rm eff}$ also plays a relevant role as it dictates the frequency of heavy oscillations about the flat minima offered by the scalar potential in charge of producing inflation. The fact that turns can happen suddenly without violating slow-roll, but remaining within the regime of validity offered by the adiabaticity condition (\ref{adiabat-cond-speed}) allows for situations in which terms proportional to $\dot c_s$ appearing in the effective theory may become relevant, and cannot be disregarded~\cite{Cespedes:2012hu}. This is precisely what happens in (\ref{eft-action-pi-inflation}), where couplings discarded in (\ref{EFT-Cheung}) do indeed appear\footnote{The result of the terms proportional to $\dot c_s$ will be contributions of new shapes to the bispectrum. We note in this regard, the observations of Ref.~\cite{raquel} in the context of general non-canonical models.}. We will discuss the consequences of this observation for the primordial bispectrum in a future report.

\section{Conclusions}
\label{sec:Conclusions}
\setcounter{equation}{0}

The effective field theory approach is a powerful organizing principle whose application to inflation~\cite{Cheung:2007st,Weinberg:2008hq} is expected to become an important tool in the comparison of theoretical models with observation.  In this paper we have attempted to clarify the non-perturbative structure of effective field theories resulting from the integration of heavy fields in multi-field models with a single light field, the inflaton, and a large mass hierarchy.

In the particularly simple case of two-field models, we have found that the
couplings of the effective field theory are given entirely in terms of the
background solution for the speed of sound $c_s(t)$ as a function of
time (dependent on the background trajectory $\phi^a_0(t)$). This result holds to all orders in the perturbative expansion and is inheritance of the time translation invariance of the parent theory, passed down to the effective theory. As might be expected, it results in relations between the $n$-point functions that furthermore, are distinguishable from other models of effective single field inflation, for example DBI inflation.

Although the parent theory we analysed is a two-field model with canonical
kinetic terms, our approach is fully covariant and we foresee no difficulty in extending the results to arbitrary sigma model metrics in field space~\cite{Gong:2011uw}. Other generalizations are expected to be more involved. In the case of multi-field models where more than one heavy field is integrated out, we might expect the appearance of other mass scales in the EFT. For instance, the iterative relations (\ref{eq: M2-M3 multi-field}) would not be so simple. 
We would like to stress that some the main features of the derived EFT, such as the emergence of a reduced speed of sound and the timescales involved can be understood in a purely Minkowski context, where the treatment of decoupling and the integration of the heavy degrees of freedom is much more tractable. Therefore, departures to any of the aforementioned simplifications assumed in this work may first be addressed in Minkowski spacetimes.

We are particularly interested in the regime of validity of the EFT and
its ability to describe departures from the usual notions of adiabaticity that were formulated for effective theories in time independent contexts. In this respect, it is useful to distinguish between ``strong'' turns, with a reduced speed of sound that varies relatively slowly (for instance, due to a fast but constant rate of turning) and ``sudden'' turns with a fast changing speed of sound (as in a sharp, sudden turn in the
trajectory). In the case of strong turns, where the speed of sound is suppressed, we find that previous arguments constraining the validity of the EFT theory apply~\cite{Baumann:2011su}, and the EFT remains valid and accurate as long as $M_{\rm eff}^2 \gg H^2$.

Sudden turns on the other hand, present another aspect of the departure from adiabaticity that may render the EFT invalid~\cite{Shiu:2011qw} in particular limits. For this case, the approach from the parent theory allows for a detailed understanding of the general adiabaticity conditions that underly the validity of the EFT, and are in fact less restrictive than has been previously assumed~\cite{Cespedes:2012hu}. When these generalized adiabatic conditions are met, the EFT is corrected by terms involving the rate of change of the speed of sound. These appear in the perturbative expansion at cubic order and were given in Section~\ref{sec:EFT}.

The generalized EFT presented here reduces to that of Ref.~\cite{Cheung:2007st} for a nearly constant speed of sound and incorporates the dominant corrections due to a sudden turn. Our computation extends the quadratic order analysis of Ref.~\cite{Achucarro:2010da,Achucarro:2010jv,Shiu:2011qw}
to all orders in the perturbations. The limit of strong but constant turns gives a concrete realization of the arguments presented in Ref.~\cite{Baumann:2011su} for the breakdown of the EFT when the speed of sound is strongly reduced, signaling new physics (which in the present case translates in the excitation of heavy modes). It also allows a controlled investigation of the effects of variations in the speed of sound~\cite{reducedcs}, as well as controlled setting for realizing models of inflation where the inflaton trajectory undergoes frequent turns~\cite{tye-xu}.

\subsection*{Acknowledgments}

We would like to thank Vicente Atal, Daniel Baumann, Cliff Burgess, Sebasti\'an C\'espedes, Cristiano Germani, Rhiannon Gwyn, Mark Jackson, Mairi Sakellariadou, Koenraad Schalm, Gary Shiu, Spyros Sypsas, Krzysztof Turzy\'{n}ski, Alexander Vikman and Jiajun Xu, for useful and interesting discussions that helped to shape this article.
GAP wishes to thank King's College London, University of Cambridge (DAMTP), CPHT at the Ecole Polytechnique for their hospitalities during the preparation of the manuscript, for which JG, GAP and SP also wish to thank Instituut-Lorentz for Theoretical Physics.
This work was supported by funds from the Netherlands Foundation for Fundamental Research on Matter (F.O.M), Basque Government grant IT559-10, the 
Spanish Ministry of Science and Technology grant FPA2009-10612 and Consolider-ingenio programme CDS2007-00042 (AA), Korean-CERN fellowship (JG), Conicyt under the Fondecyt initiation on research  project 11090279 (GAP) and the CEFIPRA/IFCPAR project 4104-2 and ERC Advanced Investigator Grants no. 226371 ``Mass Hierarchy and Particle Physics at the TeV Scale'' (MassTeV) (SP).

\end{document}